\documentstyle[twocolumn,prc,aps,epsfig]{revtex}
\input epsf
\newcommand{\be}{\begin{eqnarray}}
\newcommand{\ee}{\end{eqnarray}}

\begin{document}

\twocolumn[\hsize\textwidth\columnwidth\hsize\csname @twocolumnfalse\endcsname

\title{Rethinking the Properties of the Quark-Gluon Plasma at $T\sim T_c$} 

\author {Edward V. Shuryak and Ismail Zahed}
\address { Department of Physics and Astronomy\\ State University of New York,
     Stony Brook, NY 11794-3800}

\date{\today}
\maketitle
\begin{abstract}
We argue that although at asymptotically high temperatures
the QGP in bulk behaves as a gas of weakly interacting quasiparticles (modulo
long-range magnetism), at temperatures up to few times the critical 
temperature $T_c$ it displays different properties.
If the running of the QCD coupling constant continues in the Coulomb
phase till the screening length scale, it reaches the strong coupling treshold
$\alpha_s(m_D)\sim 1$. As a result, the Coulomb phase supports weakly bound Coulombic
s-wave $\bar c c$, light quark  and even $gg$  states.
 The existence of shallow bound states 
 dramatically increases the quasiparticle rescattering at low
 energies, reducing viscosity and
 thereby
explaining why heavy ion collisions at RHIC exhibit robust collective
phenomena. In conformal
gauge theories at finite temperature the Coulomb binding 
persists further in the strong coupling regime, as found 
for ${\cal N}=4$ SUSY YM in the Maldacena regime.
\end{abstract}
\vspace{0.1in}
]
\begin{narrowtext}
\newpage


Soon after the discovery of QCD,  it has been found that at high
temperature $T$ the color charge is $screened$ \cite{Shu_JETP} (rather than
anti-screened in the vacuum), and the corresponding phase of matter was
named the Quark Gluon Plasma (QGP). Also, it has been shown, both analytically and  
on the lattice, that at high temperature bulk quarks and gluons exist in it as
relatively free propagating quasiparticles, modulo color-magnetic
effects that are known to be non-perturbative~\cite{LINDE}.
Although explicit perturbative series for
thermodynamical quantities were found to be badly divergent, it was
still hoped that some kind of re-summation will make the weak-coupling 
quasiparticle picture work, as the screening would keep the effective coupling
weak anywhere at $T>T_c$.
The earliest suggested QGP signal was
a disappearance of familiar hadronic peaks -- $\rho,\omega,\phi$ mesons --
in the dilepton spectra \cite{Shu_QGP}. Moreover, even small-size deeply-bound
$\bar c c$ states,  $\eta_c,J/\psi$, were expected to melt at
$T>T_c$ \cite{MS,KMS}. 
 
In this paper we will show that at $T\sim T_c$ it is not the case.
 The electric effects are still strong, causing perturbation theory
to fail. But before we go into our 
reasoning, let us point out that 
a motivation  comes 
from two recent developments, the  experimental discovery of robust 
collective effects  at RHIC, known as radial and elliptic flows,
as well as recent advances in lattice simulations at $T=(1-3) T_c$.
They have led to two related questions we are going to address:
{\bf i.\,}{\em Are there  bound states of quasiparticles
 above the QCD phase transition?}
{\bf ii.\,}{\em Why does the Quark-Gluon Plasma behave like a good liquid
rather than a dilute gas of quasiparticles?}

The former question should not be confused either with 
the non-perturbative  magnetic
interaction\cite{LINDE} or with the issue of ``screening masses'' for
space-like correlators\cite{DETAR}. For earlier discussion of real
bound states above $T_c$ in the framework of high temperature QCD
sum rules see~\cite{SUM}.  Based on chiral dynamics 
driven by instanton it was found that the pi-sigma chiral multiplet 
should still exist as a resonance in the QGP \cite{Schafer:1995df}.

Recent numerical lattice studies have  found that, contrary to
earlier expectations, 
charmonium states remain  bound at least up to $T=1.5\, T_c$ \cite{Karsch_charmonium}.   
Moreover, the time-direction correlators at finite $T$ found significant
deviations
from free behavior at  $T\sim 3 T_c$  \cite{Karsch:2002wv},
in quantitative agreement with \cite{Schafer:1995df}. Their
analysis by the minimal entropy
method have suggested that pseudoscalars and even vector  resonances
exist above $T_c$. Below, we will show why these states  exist, and not just for
quarks but for gluonic plasmons as well. 

{\bf The main idea}~\cite{Shu_how} is that 
after deconfinement and chiral symmetry
restoration at $T_c$, nothing prevents  the QCD coupling from running
to 
larger values
at lower momentum scale until it is stopped at the screening
mass scale. Although at high $T$ the coupling is always small~\cite{Shu_JETP}, 
at $T\approx T_c$ 
it can reach large value $\alpha_s\sim 1$. 
As we show below, this 
 makes  Coulomb binding of the pertinent s-wave levels possible,
in spite of screening. 

  Let us start with charmonium problem first, using
the non-relativistic 
Schrodinger equation (and ignoring collisional broadening).  
For the standard radial wave function
$\chi(r)=\psi/r$  it has the usual form
with the reduced mass $m=m_c/2$.
We will use throughout this work the same in-QGP
screened Coulomb potential
of the Debye form~\cite{com1}
$ V=-(4\alpha_s(r)/3r) exp(-M_D r) $
and use the ($T$-dependent) screening mass    
as a definition of the length unit  to have its shape fixed for fixed coupling.
 In such units the equation to be discussed reads 
\be {d^2 \chi \over dx^2}+\left(\kappa^2+{4m_c\over 3M_D} {\alpha_s(x)\over
  x}e^{-x} \right)\chi=0\ee
with $\kappa^2=m_c E/M_D^2$. 
The appearance of a bound level corresponds to zero binding $E=0,\kappa=0$. If
the coupling constant $\alpha_s$ does not run and is a
constant, all parameters of the problem
appear in a single combination. Solving the equation, one can find
 the condition for the  bound state to exist 
\be \label{eqn_comb}
  {4m_c\over 3M_D} \alpha_s > 1.68\label{cond}\ee
For example,  using $4/3\alpha_s=0.471$
from the vacuum charmonium potential and $m_c=1.32 \, {\rm GeV}$, as
 Karsch et al \cite{KMS} did long ago,
 one finds a restriction on the screening  mass
$M_D< M_D^{crit}=0.37\, {\rm GeV}$. 
Lattice studies of the screening masses have been carried in
\cite{Karsch_screening} and found that for the near-critical QGP
 $M_D=(2.25\pm .25)T$. So, for $T\approx
T_c=170\, MeV$ one expects $M_D=0.34-0.42\, {\rm GeV}$. So, the condition
(\ref{cond}) is satisfied marginally if at all, and these authors 
concluded that  $\eta_c,J/\psi$  may hardly exist inside the QGP phase.

The loophole in this traditional argument is the assumption that
the gauge coupling constant remains frozen in the QGP, at the {\it same} value
it had in the in-vacuum charmonium potential. 
Our main idea is that it does not have to be so: in the QGP,
after deconfinement and chiral symmetry restoration, non-perturbative
effects are smaller and  the charge {\em continues to run to larger values}.
In Fig.\ref{fig_g_binding}(a) we compare two such potentials: although
the version with the running coupling leads to a smaller potential at
smaller distances, it is larger at larger $r$. 
One cannot of course rely on perturbative formulae in a strong coupling
regime (the alternative will be discussed elsewhere~\cite{SZ_CFT}), in this work we 
have decided to simply freeze $\alpha_s$ when it reaches 1. The cusp
in the potential reflects on that, and it occurs at about the
screening length $r\,M_D\approx 1$.

The modified potential with the running coupling leads to a more liberal 
condition for  charmonium binding, namely  $M_D<0.62\, {\rm GeV}$. This
translates into charmonium remaining bound at 
\be T<T_{\bar c c} \approx 1.6\, T_c. \ee
This value  agrees well
with recent lattice measurements \cite{Karsch_charmonium}
mentioned in the introduction,  so the
modified color potential had passed its first test.

{\bf Are there bound states of $\bar q q$, 
  and $gg$? }
For simplicity, in this work we ignore quark masses, as well as
instantons and the $U(1)$ axial anomaly, so one 
can view the chirality of light fermions to be permanently conserved
in the QGP. Chiral symmetry excludes the usual mass from being developed, and
$L,R$-handed quarks propagate independently. Nevertheless, the
propagating quark modes in the QGP have some dispersion curves with the
so called chiral mass, 
defined as the energy of the mode at zero momentum $M_q=\omega(\vec
 p=0)$, perturbatively $M_q=gT/\sqrt{6}$ to the lowest order \cite{KW}.
In weak coupling  there are two fermionic modes: {\bf i.} with the {\em same
chirality and helicity} the dispersion curve at small $p$ is
$\omega=M_q+p/3+p^2/3M_q+...$; {\bf ii.} with the opposite chirality and
helicity the mode is often called a  ``plasmino'', its dispersion curve
has a shallow minimum at $p=0.17gT$ with the energy $E_{\rm min}=0.38\,gT$
slightly below $M_q$. The latter is related to a new in-matter gluonic mode, the
plasmon~\cite{Shu_JETP}.

For a general analysis of these modes see \cite{Wel_99}, where in
particular both modes should have the same effective mass. Lattice data on the
dispersion curves for quasiparticles remains rather crude, and
restricted to essentially Coulomb gauge~\cite{latt_quasipart_masses}.
Here we will assume that $\omega^2=p^2+M^2$, with the following values (at $T=1.5 T_c$) 
\be  \label{eqn_masses}
{m_q \over T}= 3.9\pm 0.2 \hspace{1cm} {m_g \over T}= 3.4\pm 0.3\ee
Note that at such $T$ masses are large and
 their ratio  is 
very different from the weak coupling prediction $1/\sqrt{6}$, while
at $T\sim 3T_c$ and higher they are reduced toward perturbative values
(which we would not include in this work, for simplicity).

If the dispersion law is known,
the effective equation of motion suitable for discussion
of the bound state problem can be obtained by
standard substitution of the covariant
derivatives in the place of momentum and frequency. It has the form of
the Schrodinger equation if
the  dispersion curve can be parameterized as
$\omega=M_q+p^2/2M'+..$, in general with two different
constants $M,M'$. Both for weak coupling and lattice data,
such approximation seem to be accurate    
withing several percents.

Now, to address the issue of binding,
we first note that if all effective masses  grow linearly with $T$,
including the screening mass, the explicit $T$ dependence drops out
of (\ref{eqn_comb}) to the exception of the logarithmic
dependence in $T$ left out in $\alpha_s\sim 1/{\rm ln}(T/\Lambda_{QCD})$.
This is why the region of ``strongly coupled QGP'' turns out to be 
relatively substantial. For a qualitative estimate, let us
set the  coupling to its maximum, $\alpha_s=1$. The combination of 
constants is $(4\times 3.9\,T)/(3\times 2.25\,T)=2.3$, larger than the
critical value (\ref{eqn_comb}), so one should expect the 
occurrence of (strong) Coulomb bound states. Although the (plasmon)
gluon modes are somewhat lighter than quarks in (\ref{eqn_masses}),
their Coulomb interaction has a larger coefficient due to
a different Casimir operator for the adjoint representation, 3
instead of 4/3. As a result the effective combination in the potential is
$3 m_g \alpha_s/M_D$, which  is about twice larger than for quarks, and
thus the gluons are bound even stronger (modulo collisional broadening).

Solving the equations one can make a quantitative analysis, using the same
potential as above. We found that the highest temperature $T$ at which
light quark states are Coulomb bound is somehow lower than that of charmonium,
\be
T_{\bar q q}\approx 1.45\, T_c \approx 250 \, {\rm MeV}\,\,,\ee while the 
s-wave $gg$ gluonium states 
remain bound
till higher temperatures  
$
T_{gg}\approx 4 \,T_c $.
As an example,  in Fig.~\ref{fig_g_binding}(b) we show the 
binding energy of two gluons. (In absolute units it reaches 
 about 100 MeV, while it is much smaller MeV for quarks.)
The  wave functions of all loosely bound
 states are similar to 
 that of a deuteron, with the usual $\chi\sim e^{-\kappa r}$
behavior outside  the potential.

How many bound states are there? Stating with quarks, we note that if one
ignores current quark masses and instanton effects, the chiralities
(left and right) are conserved. Furthermore,
there are two different modes of
quarks depending on helicity (``particle'' and ``plasmino''), times 
antiquarks, times
flavors. All in all $4 N_f^2$ states, connected by continuity
to the pseudoscalar, 
scalar, vector and axial vector nonets in the vacuum 
(see Fig.\ref{fig_masses_T}). 

\begin{figure}[t]
\centering
\includegraphics[width=4cm]{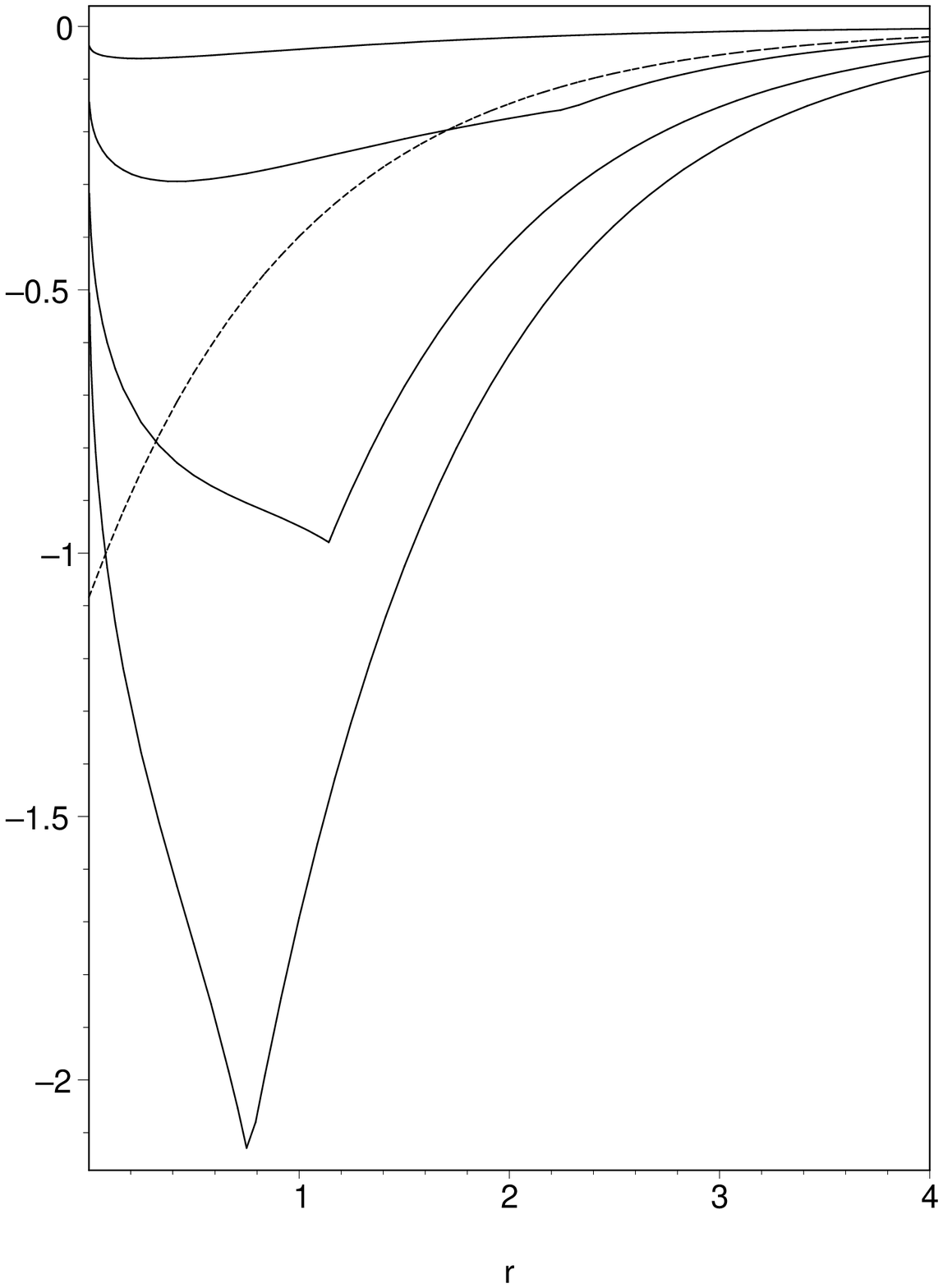}
\includegraphics[width=4.4cm]{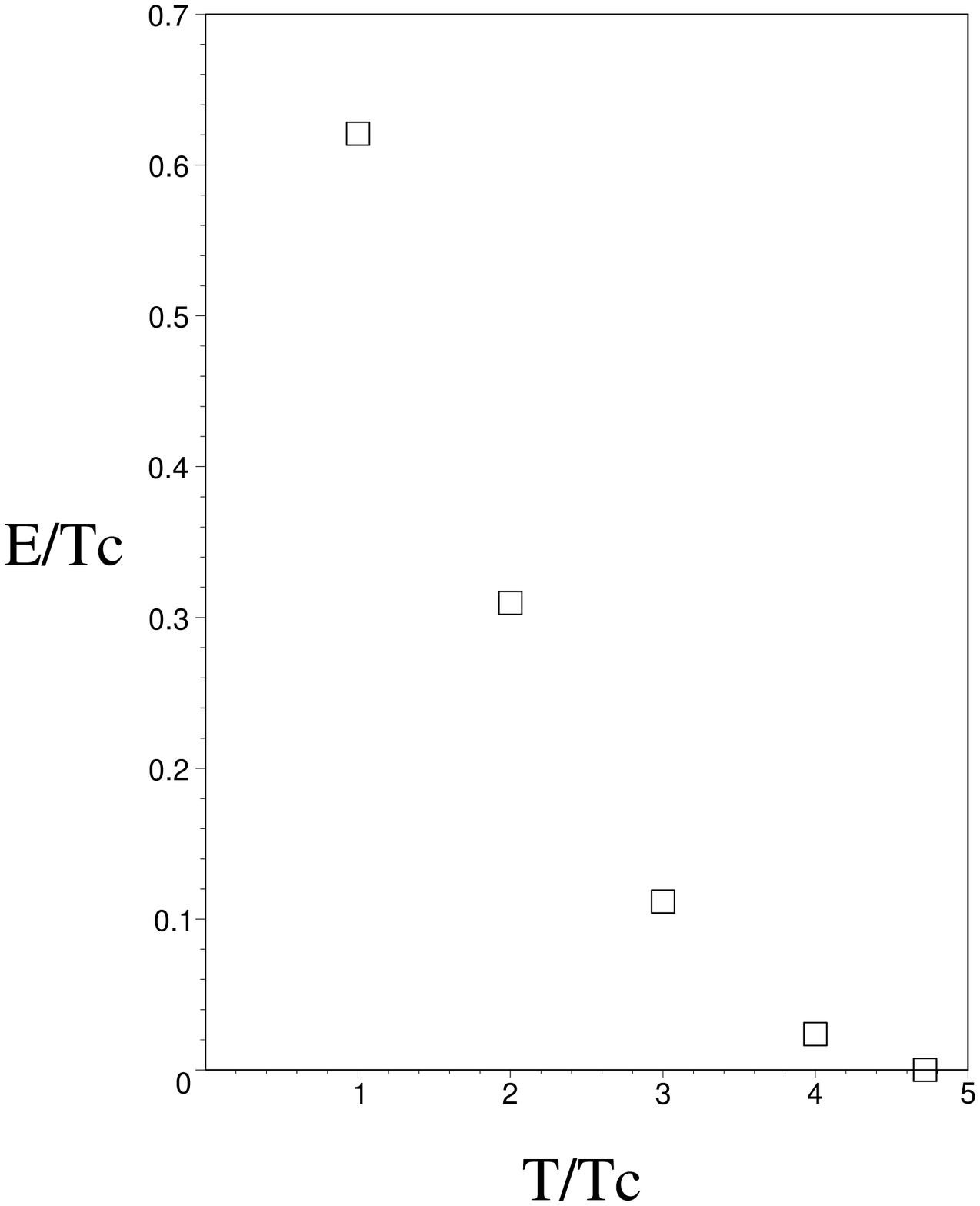}
\caption{\label{fig_g_binding} 
(a)The combination $(M_c/M_D) rV(r)$, versus 
the distance $r M_D$.
The dashed line corresponds  to  $4\alpha_s/3=0.471$, while
 the solid lines correspond to the running coupling constant,
for $T=1;1.5;3;10 T_c$, from bottom up. The cusps  
 occur when $\alpha_s$ reaches 1.
(b) The (modulus of the) binding energy of two gluons, versus $T/T_c$.
}
\end{figure}

Now we come to gluons: both  may have any of  3 polarizations (
with longitudinal ``plasmons in matter). Furthermore (as discussed 
in our longer paper \cite{SZ_CFT}) there can be binary composites even
if the total charge is non-zero, which provides  $\sim N_c^2$ states.
 On top of that,
 one can also think of hybrid (qg)
composites, with $\sim N_f N_c$ states. These exotic states
however need stronger coupling and may exist only very close to $T_c$,
if at all.

{\bf  QGP kinetics} was studied in weak coupling,
with a rather large mean free path,  $T\,l_{\rm mfp}\sim
1/g^4 {\rm ln}\,(1/g) \gg 1$. Similar pQCD-inspired ideas have
led to the somehow pessimistic expectation that the Relativistic 
Heavy Ion Collider (RHIC) project in Brookhaven National Laboratory, 
would produce a firework of multiple jets and mini-jets with 
perturbative corrections for rescattering. To the contrary,
the very first RHIC  run in the summer of 2000 has shown
spectacular collective motion patterns, known as
radial and elliptic flows. The spectra of about 99\%
of all kinds of secondaries, except the high-$p_t$ tails, were shown to be
described by hydrodynamics~\cite{hydro}, based on the opposite -- zero mean free
path  -- regime. Further studies of partonic cascades~\cite{GM} 
and viscosity corrections~\cite{Teaney_visc} 
have found that the data can only be explained if 
partonic cross sections are increased by huge factors, from 2-3 mb
for pQCD estimates to about 100 mb. The equation of state used 
in~\cite{hydro} is not very different from lattice predictions and 
even simple weak coupling expectations.

A sequence of loosely bound 
 states (black dots in  Fig.\ref{fig_masses_T})
should be very important for the QGP kinetics.
The scattering cross section may be strongly enhanced by low-lying
resonances, as known examples from nuclear physics (such as
low-energy neutron-capture resonances) clearly demonstrate.
The Wigner cross section of slow particles
\be \sigma(E)= {2\pi \over m} {1\over E+|E_{\rm b} | }\ee
has a peak when there is both a shallow level and a virtual state,
as we know from NN scattering in the triplet and singlet channels\cite{com3}.
 With the denominator   $E+|E_{\rm b}|$
ranging from 100 to 10 MeV, the corresponding $gg$ cross section is
in the range 50-500 mb. These huge cross sections make
quasi-free propagation of quasiparticles impossible.

We stress that our analysis is not based  on the in-vacuum  
effective  potential~\cite{com4}, but
on the  in-matter screened one. The bound states emerge in 
a screened but still strongly coupled Coulomb phase.
Still, not all of the many-body
effects are included in the screening. Indeed, the range
of distances in the  weakly bound states is large,
and by including the $1/N_c^2$ factor needed for absolute
matching of the gluon colors, we  may even have {\em more than one}
matching gluon inside the available bound-state volume.
More detailed many-body studies should be made, such as the ones 
carried for ``excitonic matter'' in semiconductors and insulators.
We recall that the Coulomb parameter in QED plasmas is
$e^2/\epsilon\, v$ where $e^2=1/137$ is the usual fine structure constant,
$\epsilon$ is the dielectric constant of the background substance and
$v$ is a typical particle/hole velocity. It can in principle be tuned to be
around 1, resulting in bound excitons
(particle-hole pairs). The latter occur despite
Debye-screening, a situation much alike ours. Depending on a number of
parameters, including the density of excitons and temperature, 
the system exhibits various phases, ranging from an ideal gas of excitons to
a liquid or plasma, or even a Bose-condensed gas. On its way from a gas
to a liquid, clustering with 3- and 4-body states play an important role.
Although one cannot directly relate these two problems
(quarks and gluons have $N_c$ and $(N_c^2-1)$ colors respectively, while particles
and holes have simply charges $\pm e$), one may think that in the QGP at $T\sim
T_c$ some of these phenomena may well be there. 

More generally,  strongly coupled many-body problems,
 with large or divergent
 binary scattering length, are studied in 
at least two other settings: (i) a  gas of neutrons, with $a=18 \, fm$
due to a virtual level;  (ii)
trapped atomic $Li^6,Li^7$ in which the scattering length can be tuned. 
Its hydrodynamic properties are of great interest.

{\bf Discussion and  Outlook.}
The main assumption made in this work is that at $T>T_c$ the gauge
coupling
is allowed to run
till the screening scale.
 As a result, we found that the QCD phase above $T_c$ supports
Coulomb bound states. In this work we were mainly interested in the
location of the point at which the s-wave state disappear. However,
more generally, one may ask how these states are related to know
hadrons
at $T=0$. How it supposed to happen is shown in Fig.\ref{fig_masses_T}
(for 2 massless flavors, when the phase transition is second order
and all lines must be continuous). Very close to
 $T_c$ the quark (chiral) mass is expected to become large, so here
mesons are rather deeply bound 
states in this region and 
both relativistic and non-perturbative effects become important: we
plan to study this region elsewhere.
 It would also be interesting to study whether
annihilation of vector states can be seen in the dilepton spectra,
and how the survival of charmonium above $T_c$ affects its evolution
at RHIC.

\begin{figure}
\centering
\includegraphics[width=7.cm]{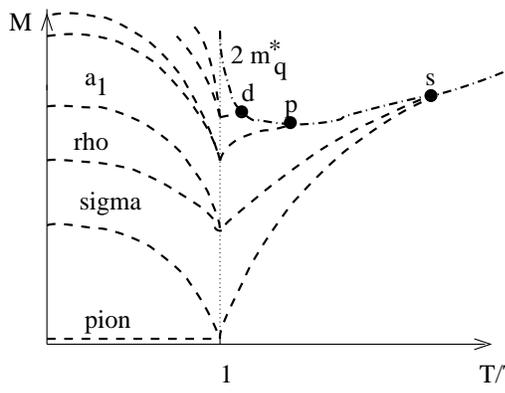}
   \caption{\label{fig_masses_T}
Schematic dependence of hadronic masses on temperature $T$ (in units
of the critical one $T_c$), for 2-flavor QCD in the chiral limit.
The dash-dotted line corresponds to twice the (chiral) effective mass
of a quark. Black dots marked $s,p,d$ correspond to the points where
the binding vanishes for
states with orbital momentum $l=0,1,2...$.
}
\end{figure}

In QCD, as soon as the lowest state hits zero there is a phase
transition. This is however impossible for 
 conformal gauge theories (CFTs), such as $\cal N$=4 supersymmetric
 gauge
 theory. In CFT 
the gauge coupling is allowed to become {\em supercritical} or even large 
 $\lambda\equiv g^2 N_c \gg 1$. 
 Maldacena's AdS/CFT duality 
\cite{Maldacena} has opened a way to study this {\em strong coupling limit} 
using classical gravity. At finite $T$ it was recently
actively discussed, for Debye screening \cite{Rey_etal},
bulk  thermodynamics \cite{thermo} and kinetics \cite{PSS}.
Although the thermodynamical quantities are only modified by an
overall factor of 3/4 in comparison to the black-body limit,
kinetics is changed dramatically.
 In our separate paper 
\cite{SZ_CFT} we  show that in this regime the matter
is made of very deeply bound binary composites, in which the supercritical
Coulomb can be balanced by centrifugal force.
a re-summation of a class of diagrams, in vacuum and at finite $T$. 
Specific towers of such bound states can be considered as a
continuation
of Fig.\ref{fig_masses_T} to the left, toward stronger and stronger coupling. 

{\bf Acknowledgments.}
We thank Gerry Brown for many valuable discussions.
This work was supported in parts by the US-DOE grant DE-FG-88ER40388.

\end{narrowtext}

\begin{thebibliography}{99}
\bibitem{KMS}
F.~Karsch, M.~T.~Mehr and H.~Satz,
Z.\ Phys.\ {\bf C37} (1988)  617.
(Note that this work included in the potential 
a screened version of the confining
potential, for continuity with the $T=0$ limit, which we dont need.
\bibitem{com1} At strong coupling in conformal gauge theory
 we find a slightly different shape for the screening
\bibitem{com3}We remind that the deuteron is bound by
$2.2 \, MeV$, while the singlet-isosinglet  virtual level
is at $67 \,{\rm keV}$.
\bibitem{com4} These shallow states
affect the propagation of slow neutrons in hydrogen-containing matter, but not
in dense nuclear matter where these levels and the
associated cross section disappear due to in-matter screening.
\end{thebibliography}
\end{document}